\input harvmac\skip0=\baselineskip
\input epsf
\noblackbox
\newcount\figno
\figno=0
\def\fig#1#2#3{
\par\begingroup\parindent=0pt\leftskip=1cm\rightskip=1cm\parindent=0pt
\baselineskip=11pt
\global\advance\figno by 1
\midinsert
\epsfxsize=#3
\centerline{\epsfbox{#2}}
\vskip 12pt
{\bf Fig.\ \the\figno: } #1\par
\endinsert\endgroup\par
}
\def\figlabel#1{\xdef#1{\the\figno}}


\lref\Gaberdiel{ M.~R.~Gaberdiel and P.~Kaste,
``Generalised discrete torsion and mirror symmetry for G(2) manifolds,''
arXiv:hep-th/0401125.
}
\lref\joyce{D.~D.~Joyce,
``Compact 8-manifolds with holonomy Spin(7),''
Inv.\ Math.\ 123 (1996) 507.
}
\lref\joycegtwo{D.~D.~Joyce, ``Compact Riemannian 7-manifolds with
holonomy $G_2$. I,'' J.\ Diff.\ Geom.\  43 (1996) 291}
\lref\joycegtwotwo{ D.~D.~Joyce, ``Compact Riemannian 7-manifolds
with holonomy $G_2$. II,'' J.\ Diff.\ Geom.\  43 (1996) 329. }
\lref\majumder{ J.~Majumder,
``Type IIA orientifold limit of M-theory on compact Joyce 8-manifold of
Spin(7)-holonomy,''
JHEP {\bf 0201}, 048 (2002) [arXiv:hep-th/0109076].
}
\lref\lerche{ W.~Lerche, C.~Vafa and N.~P.~Warner,
``Chiral Rings In N=2 Superconformal Theories,''
Nucl.\ Phys.\ B {\bf 324}, 427 (1989).
}
\lref\SYZ{ A.~Strominger, S.~T.~Yau and E.~Zaslow,
``Mirror symmetry is T-duality,''
Nucl.\ Phys.\ B {\bf 479}, 243 (1996) [arXiv:hep-th/9606040].
}
\lref\VW{ C.~Vafa and E.~Witten,
``On orbifolds with discrete torsion,''
J.\ Geom.\ Phys.\  {\bf 15}, 189 (1995) [arXiv:hep-th/9409188].
}
\lref\acharya{ B.~S.~Acharya,
``On mirror symmetry for manifolds of exceptional holonomy,''
Nucl.\ Phys.\ B {\bf 524}, 269 (1998) [arXiv:hep-th/9707186].
}
\lref\acharyatwo{B.~S.~Acharya, ``Exceptional mirror symmetry''
Winter School on Mirror Symmetry, Vector Bundles and Lagrangian
Submanifolds (Cambridge, MA, 1999), P.1--P.14, AMS/IP Stud. Adv.
Math., 23. }
\lref\acharyathree{B.~S.~Acharya, ``Dirichlet Joyce manifolds,
discrete torsion and duality,'' Nucl.\ Phys.\ B {\bf 492}, 591
(1997) [arXiv:hep-th/9611036].
}
\lref\acharyafour{ B.~S.~Acharya and S.~Gukov, ``M Theory And
Singularities Of Exceptional Holonomy Manifolds,'' Phys.\ Rept.\
{\bf 392}, 121 (2004).
}
\lref\acharyafive{B.~S.~Acharya, ``N=1 M-theory-Heterotic Duality
in Three Dimensions and Joyce Manifolds,'' arXiv:hep-th/9604133.
}

 \lref\Farrill{ J.~M.~Figueroa-O'Farrill, ``A
note on the extended superconformal algebras associated with
manifolds of exceptional holonomy,'' Phys.\ Lett.\ B {\bf 392}, 77
(1997) [arXiv:hep-th/9609113].
}
\lref\keshav{ M.~Becker, K.~Dasgupta, A.~Knauf and R.~Tatar,
``Geometric transitions, flops and non-Kaehler manifolds. I,''
arXiv:hep-th/0403288.
}
\lref\Gurrieri{ S.~Gurrieri, J.~Louis, A.~Micu and D.~Waldram,
``Mirror symmetry in generalized Calabi-Yau compactifications,''
Nucl.\ Phys.\ B {\bf 654}, 61 (2003) [arXiv:hep-th/0211102].
}
\lref\gukov{ S.~Gukov, J.~Sparks and D.~Tong,
``Conifold transitions and five-brane condensation in M-theory on Spin(7)
manifolds,''
Class.\ Quant.\ Grav.\  {\bf 20}, 665 (2003)
[arXiv:hep-th/0207244].
}
\lref\cvetic{ M.~Cvetic, G.~W.~Gibbons, H.~Lu and C.~N.~Pope,
``Cohomogeneity one manifolds of Spin(7) and G(2) holonomy,''
Phys.\ Rev.\ D {\bf 65}, 106004 (2002) [arXiv:hep-th/0108245].
}
\lref\SV{ S.~L.~Shatashvili and C.~Vafa,
``Superstrings And Manifold Of Exceptional Holonomy,''
arXiv:hep-th/9407025.
}
\lref\book{D.~A.~Cox and S.~Katz, ``Mirror Symmetry And Algebraic
Geometry,'' American Mathematical Society, (September 1, 1999)
}
\lref\candeles{P.~Candelas, X.~C.~De La Ossa, P.~S.~Green and L.~Parkes,
``A Pair Of Calabi-Yau Manifolds As An Exactly Soluble Superconformal
Theory,''
Nucl.\ Phys.\ B {\bf 359}, 21 (1991).
}
\lref\odake{ S.~Odake, ``Extension Of N=2 Superconformal Algebra
And Calabi-Yau Compactification,'' Mod.\ Phys.\ Lett.\ A {\bf 4},
557 (1989).
}
\lref\VafaTorsion{C.~Vafa, ``Modular Invariance And Discrete
Torsion On Orbifolds,'' Nucl.\ Phys.\ B {\bf 273}, 592 (1986).
}

\Title{\vbox{\baselineskip12pt\hbox{SLAC-PUB-10504}\hbox{SU-ITP-04/25}
}} {\vbox{\vskip -1.5cm\centerline{A Note on Mirror Symmetry for
Manifolds} \centerline{with Spin(7) Holonomy}}}

\centerline{Wu-yen Chuang}
\centerline{SLAC, and Department of Physics} \centerline{Stanford
University, Stanford, CA 94305} \centerline{wychuang@itp.stanford.edu}

\vskip .3in \centerline{\bf Abstract} { Starting from the
superconformal algebras associated with $G_2$ manifolds, I extend
the algebra to the manifolds with spin(7) holonomy. I show how the
mirror symmetry in manifolds with spin(7) holonomy arises as the
automorphism in the extended sperconformal algebra. The
automorphism is realized as 14 kinds of T-dualities on the
supersymmetric $T^4$ toroidal fibrations. One class of Joyce's
orbifolds are pairwise identified under the symmetry. }

\smallskip
\Date{}
\newsec{Introduction}
Mirror symmetry is a beautiful subject both in physics and
mathematics. It was first conjectured in \lerche \  that there
exists a symmetry which exchanges the complex moduli on one
manifold with the Kahler moduli on the dual manifold when we
consider the string worldsheet propagation on Calabi-Yau target
spaces. The symmetry arises in the sense that the resulting
physical spectra of the mirror pair are isomorphic. This requires
the Betti numbers of the CY mirror pair satisfy the condition
$b_{p,q}(M)=b_{d-p,q}( \tilde{M})$. It was also shown that mirror
symmetry could determine non-perturbative effects of worldsheet
instantons by counting the number of holomorphic curves in
Calabi-Yau spaces \candeles . Those who are interested in various
aspects of mirror symmetry are referred to \book .

In \SYZ, Strominger, Yau and Zaslow (SYZ) argued that the mirror
transformation is equivalent to T-duality on the supersymmetric
$T^3$ fibration in the Calabi-Yau manifolds, by considering the
mirror BPS soliton spectra in two theories (IIA/IIB). Some
concrete mirror pairs of certain toroidal orbifolds with discrete
torsion can be found in \VW, where the mirror symmetry is indeed
realized as T-duality on toroidal $T^3$ fibration in the
orbifolds. And these examples involving the changes of discrete
torsion are related to the main goal of this paper.

In \acharya \acharyatwo, Acharya discussed the existence of the
mirror symmetry in IIA/IIB string theory compactified on manifolds
with exceptional holonomy and argued how the discrete torsion
trasforms under the $T^4$ T-duality. In \Gaberdiel, the authors
gave some concrete mirror pairs among Joyce's orbifolds with $G_2$
holonomy, which are built from resolving or deforming $T^7/Z_2^3$
orbifolds. They also identified the mirror symmetry as an
automorphism in the extended superconformal algebra on manifolds
with $G_2$ holonomy.

Motivated by these known results, I generalize the chiral
superconformal algebra to the manifolds with spin(7) holonomy, and
identify the corresponding automorphism in the algebra as a
combination of a T-duality in 8-direction and a generalized $G_2$
mirror transformation or a combination of two distinct $G_2$
mirror transformations. The automorphism could also be understood
as T-duality on the supersymmetric $T^4$ fibrations. In order to
make the automorphism clearer, I give an example of one class of
Joyce's spin(7) manifolds with discrete torsion. The 14 kinds of
$T^4$ T-dualities are classified into two categories, one of which
does flip the discrete torsion and hence lead to a topologically
different Joyce's orbifold and the other does not.

The paper is presented as follows. In section 2 I review the
mirror symmetry of Calabi-Yau and $G_2$ manifolds both from the
viewpoint of the conformal field theory and the T-duality. In
section 3 I will give the construction of spin(7) extended
superconformal algebra and identify the automorphism in it as 14
kinds of T-dualities and classify them into two kinds as mentioned
above. Section 4 is conclusion and some suggestion for future
study.

\newsec{Mirror symmetry for CY and $G_2$ manifolds }

In this section I will give a short review of mirror symmetry on
Calabi-Yau orbifolds ($T^6/Z_2^2)$ and Joyce's $G_2$ manifolds
\Gaberdiel \joycegtwo \joycegtwotwo.

\subsec{Mirror symmetry of Calabi-Yau threefolds}

The generators of the $N=2$ superconformal algebra for string
propagation on Calabi-Yau target-space are the stress energy
tensor $T_{CY}$, two supercurrent $G_{CY}$, $G'_{CY}$ and the U(1)
current $J_{CY}$, along with a complex current $\Omega_{CY}$ of
conformal weight $3/2$ constructed from the worldsheet fermions
and its superpartner $\Psi_{CY}$.

In $T^6/Z_2^2$ orbifolds, they can be expressed as, \Gaberdiel \Farrill \odake

\eqn\CYgenerator{ \eqalign{ T_{CY} &= {1 \over 2} \sum_{j=1}^6 :
\partial x^j \partial x^j : - {1 \over 2} \sum_{j=1}^6 : \psi^j
\partial \psi^j : \ , \cr G_{CY} &= \sum_{j=1}^6 : \psi^j \partial x^j
: \ , \ G'_{CY} = \sum_{j=1}^3 ( \psi^{2j-1} \partial x^{2j}-
\psi^{2j}
\partial  x^{2j-1}) \ , \
J_{CY} = \sum_{j=1}^3 \psi^{2j-1}\psi^{2j} \ , \cr \Omega_{CY} &=
\psi^1 \psi^3 \psi^5 -  \psi^1 \psi^4 \psi^6 - \psi^2 \psi^3
\psi^6 -  \psi^2 \psi^4 \psi^5 + i( \psi^1 \psi^3\psi^6 + \psi^1
\psi^4 \psi^5 +  \psi^2 \psi^3 \psi^5 - \psi^2 \psi^4 \psi^6)\ ,
\cr \Psi_{CY} &:= \{ G_{CY}, \Omega_{CY} \} \ . }}

There exists an automorphism in the superconformal algebra or OPE,
which leave invariant the $N=1$ superconformal subalgebra generated
by $T_{CY}$ and $G_{CY}$.

\eqn\mirrorCY{G'_{CY} \to -G'_{CY}, \ \ J \to - J, \ \, \Omega \to
\Omega^{*}, \ \ \Psi \to \Psi^{*} .}

Calabi-Yau mirror symmetry is to apply the above automorphism to
one of the chiralities of the algebra, for instance,
$\tilde{G}'_{CY}$, $\tilde{J}_{CY}$, $\tilde{\Omega}_{CY}$, and
$\tilde{\Psi}_{CY}$. Recall that the T-duality in $i$th direction
will leave $\partial x_i$ and $\psi_i$ invariant but reverse
$\bar{\partial} x_i$ and $\tilde{\psi}_i$. Therefore, we can
easily see that the T-duality on $T^3$ fibrations in the following
directions ( which appear in the indices of $\Omega_{CY}$ ) also
generates the mirror symmetry.

\eqn\directions{ \{
(1,3,5),(1,4,6),(2,3,6),(2,4,5),(1,3,6),(1,4,5),(2,3,5),(2,4,6)
\}}

Some concrete examples of these T-dualities acting on the $T^3$ fibration
and changing the discrete torsion can be found in \VW \Gaberdiel.

\subsec{Compact orbifolds with $G_2$ holonomy}

In this and the following sections, I will first give an example
of Joyce's orbifolds which were constructed by disingularising
$T^7/Z_2^3$ and how the choices in resolving (deforming) the
singularities can result in topologically different spaces. After
that, I will write down the $G_2$ extended chiral superconformal
algebra and look for the automorphism in it. We will see that
applying the automorphism transformation to one of two chiralities
is equivalent to applying a T-duality on certain $T^3$ toroidal
fibration.

Consider the orbifolds of $T^7 / \Gamma$, where $x_i=x_i+1$ and
$\Gamma$ is generated by three $Z_2$,\joycegtwo

\eqn\gtwo{\eqalign{ \alpha &= ( -x_1, -x_2, -x_3, - x_4, x_5, x_6,
x_7) \ , \cr \beta &= ( -x_1, 1/2 - x_2, x_3, x_4, -x_5, -x_6,
-x_7) \ , \cr  \gamma &= ( -x_1, x_2, -x_3, x_4, -x_5, x_6, -x_7)\
. }}

In order to desingularize the orbifolds, one has to know, for
instance, how the 16 $\alpha$ fixed $T^3s$ get identified under
the group generated by $\beta$ and $\gamma$. What we found in this
example is the 16 $T^3s$ fixed by $\alpha$ or $\beta$ are reduced
to 4 orbits of order 4 by the free-acting of the $<\beta, \gamma>$
or $<\gamma, \alpha>$. In the $\gamma$-fixed $T^3$ sector, the
group $<\alpha, \beta>$ only reduce 16 $T^3$ to 8 orbits of order
2 since $\alpha\beta$ acts trivially on them.

The choices of blowing-up or deforming also come from this
$\gamma$-fixed sector. From a discrete torsion analysis based on
the requirement of modular invariance \Gaberdiel, we know that
blowing-up (deforming) corresponds to discrete torsion in the
$\gamma$-fixed sector $\epsilon_{\gamma; \tilde{f}}=1 \ (-1)$ and
the even (odd) $\alpha \beta$ parity. By virtue of the
correspondence between the RR ground states and the cohomology, we
can write down the RR ground states in $\gamma$-fixed sector.

For $\alpha \beta$ parity even case, we have, \eqn\RRstates{
\eqalign{ &\epsilon_{\gamma; \tilde{f}}=1 \ , \cr
&|0,0;\tilde{f}\rangle_{\gamma} \ , \ \psi^{2+}
|0,0;\tilde{f}\rangle_{\gamma}\ ,\
\psi^{4+}\psi^{6+}|0,0;\tilde{f}\rangle_{\gamma}\ , \
\psi^{2+}\psi^{4+}\psi^{6+}|0,0;\tilde{f}\rangle_{\gamma}\ , }}
where $\tilde{f}=1,..8$ labelling the $\gamma$-fixed points after
$\alpha$ or $\beta$ identification.

For $\alpha \beta$ parity odd case, the RR ground states are,
\eqn\RRstatestwo{ \eqalign{ &\epsilon_{\gamma; \tilde{f}}=-1\ ,
\cr &\psi^{4+}|0,0;\tilde{f}\rangle_{\gamma}\ , \
\psi^{6+}|0,0;\tilde{f}\rangle_{\gamma}\ ,\
\psi^{2+}\psi^{4+}|0,0;\tilde{f}\rangle_{\gamma}\ , \
\psi^{2+}\psi^{6+}|0,0;\tilde{f}\rangle_{\gamma}\ . }}

One should regard $|0,0;\tilde{f}\rangle_{\gamma}$ as the harmonic
two form associated with the exceptional divisors of the
blowing-up (deformation). Therefore, blowing-up contributes 1 to
$b_2$ and 1 to $b_3$ while the deformation increases $b_3$ by 2.

For the RR ground states in $\gamma$-fixed sector, the operation
$\alpha \beta$ reverses the 4th and 6th directions. Therefore, we
can express it as,

\eqn\AB{ \alpha \beta = {1 \over 4} \psi_0^4 \psi_0^6
\tilde{\psi}_0^4 \tilde{\psi}_0^6 \epsilon_{\gamma ; \tilde{f}} \
\ .}

We denote $X_l$ Joyce's manifold with $l$ blow-ups and $8-l$
deformations. After summing up all Betti numbers from various
sectors, we have,

\eqn\bettigtwo{ (b_0,...,b_7)=(1,0,8+l,47-l,47-l,8+l,0,1)\ .}

\subsec{$G_2$ extended superconformal algebra}

The algebra on manifolds with $G_2$ holonomy is generated by
appending a spin $3/2$ operator $\Phi_{G_2}$ and its superpartner
$X_{G2}$ to the $N=1$ superconformal subalgebra spanned by
$T_{G_2}$ and $G_{G_2}$ \Farrill \SV. In our basis of coordinates,
they are,

\eqn\gtwogen{ \eqalign{ T_{G_2} &= {1 \over 2} \sum_{j=1}^7 :
\partial x^j
\partial x^j : - {1 \over 2} \sum_{j=1}^7 : \psi^j  \partial
\psi^j : \ , \ \ G_{G_2} = \sum_{j=1}^6 : \psi^j \partial x^j : \
, \cr \Phi_{G_2} &= \psi^1 \psi^3 \psi^6 + \psi^1 \psi^4 \psi^5 +
\psi^2 \psi^3 \psi^5- \psi^2 \psi^4 \psi^6 +\psi^1 \psi^2 \psi^7
+\psi^3 \psi^4 \psi^7 +\psi^5 \psi^6 \psi^7 \ , \cr X_{G2} &= -
\psi^2 \psi^4 \psi^5 \psi^7 - \psi^2 \psi^3 \psi^6 \psi^7 -\psi^1
\psi^4 \psi^6 \psi^7 +\psi^1 \psi^3 \psi^5 \psi^7 - \psi^3 \psi^4
\psi^5 \psi^6 \cr \ \ \ \ &\ -\psi^1 \psi^2 \psi^5 \psi^6 - \psi^1
\psi^2 \psi^3 \psi^4 - {1 \over 2} \sum_{j=1}^7 : \psi^j
\partial \psi^j: \ . }}

The extended superconformal algebra has one obvious automorphism
\Farrill \SV.

\eqn\mirrorGtwo{ \Phi_{G2} \to -\Phi_{G_2};\ \ K_{G2} \to
-K_{G2};\ \ T_{G2} \ , \ G_{G2} \ , \  X_{G2} \ , \ M_{G2} \
unchanged.}

If the $G_2$ manifolds are of the form $(CY_3 \times S^1) / Z^2$
as the Joyce $G_2$ manifolds, we can also reformulate the
superconformal generators in terms of the Calabi-Yau ones.

\eqn\gtwogenerator{ \eqalign{ T_{G_2} &= T_{CY} + {1 \over 2} :
\partial x^7 \partial x^7 : - {1 \over 2} : \psi^7 \partial \psi^7
: \ , \ \ G_{G_2} = G_{CY} + : \psi^7 \partial x^7: \ ,  \cr
\Phi_{G_2} &= Im(\Omega_{CY}) + : J_{CY} \psi^7 : \ , \cr X_{G_2}
&= \ \  : Re(\Omega_{CY}) \psi^7: + { 1\over 2} : J_{CY}J_{CY} : -
{1 \over 2} : \psi^7\partial \psi^7: \ ,  \cr K_{G2} &= Im
(\Psi_{CY}) + : J_{CY}
\partial x^7 : + : G'_{CY} \psi^7: \ , \cr M_{G2} &= \ \ :Re (\Psi_{CY})
\psi^7:-:Re(\Omega_{CY}) \partial x^7: + :
\partial x^7 \partial \psi^7:+ :J_{CY}G'_{CY}:- {1 \over 2}
\partial G_{CY} \ . }}

Similarly, the generalized mirror symmetry for manifolds with
$G_2$ holonomy is to apply the above automorphism to one of the
two chiralities. On the other hand, the T-duality in the following
$(i_1, i_2, i_3)$ directions can obviously realize the
automorphism.

\eqn\directiongtwo{ \eqalign{  &(i_1, i_2, i_3)  \in I_3^{+} \cup
I_{3}^{-}\ , \cr &I_{3}^{+} = \{ (2,4,6) , (2,3,5) , (1,2,7) \}\ ,
\cr &I_{3}^{-} = \{ (1,3,6) ,(1,4,5) , (3,4,7) , (5,6,7) \} \ . }
}

If we combine any two different T-dualities listed above, we
obtain another set of T-dualities acting on toroidal $T^4$, which
also leave the extended chiral algebra invariant. Hence, they are
mirror symmetry which take IIA (IIB) to IIA (IIB).

\eqn\directiongtwotwo{ \eqalign{ &(i_1, i_2, i_3, i_4 ) \in
I_4^{+} \cup I_4^{-}\ ,  \cr  &I_{4}^{+} = \{
(1,3,5,7),(1,4,6,7),(3,4,5,6) \}\ , \cr &I_4^{-} = \{
(2,4,5,7),(2,3,6,7),(1,2,5,6),(1,2,3,4) \}\ . }}

Recall that T-duality in $i$th direction will give
$\tilde{\psi}_0^{i}$ a minus sign. It's not hard to see that
$I_3^{+}$ ( $I_4^{+}$ ) does not change the discrete torsion while
$I_3^{-}$ ( $I_4^{-}$ ) does. We can summarize the action of the
T-dualities as follows.

\eqn\gtwosummary{ \eqalign{  IIA(IIB) / X_l &\longleftrightarrow
IIB(IIA)/ X_{8-l}\ , \ \ under \ I_3^{-} \ , \cr IIA(IIB) / X_l
&\longleftrightarrow IIA(IIB)/ X_{l}\ , \ \  under \ I_3^{+}\ .}}


\newsec{Mirror symmetry for spin(7) manifolds}

\subsec{Joyce's construction of spin(7) manifolds}

There are many known examples of Joyce spin(7) orbifolds \joyce .
For simplicity, I will take one orbifold for example in which we
have choices in desingularizing $T^8/Z_2^4$ as before. The
generators are,

\eqn\spin{ \eqalign{ \alpha &= ( -x_1, -x_2, -x_3, - x_4, x_5,
x_6, x_7, x_8)\ ,  \cr \beta &= ( x_1, x_2, x_3, x_4, -x_5, -x_6,
-x_7, -x_8)\ ,  \cr \gamma &= ( 1/2-x_1, -x_2, x_3, x_4, 1/2-x_5,
-x_6, x_7, x_8)\ ,  \cr \delta &= ( -x_1, x_2, 1/2-x_3, x_4,
1/2-x_5, x_6, 1/2-x_7, x_8)\ .}}

Again, the periodicity of $x_i$ is unity. In general, the
singularities arises in five different types and the corresponding
desingularization is following.

Type(1): increase $b_2$ by 1, $b_3$ by 4, $b_{4+}$ by 3, and
$b_{4-}$ by 3. The singularity type is $T^4 \times ( B_{\epsilon}^4 / \{ \pm1 \})$, where
$B_{\epsilon}^4$ is defined as an open ball of radius $\epsilon$ about $0$ in $R^4$.

Type(2): increase $b_2$ by 1, $b_{4+}$ by 3, and $b_{4-}$ by 3.
The singularity if of the form $( T^4 / \{ \pm1 \} \times ( B_{\epsilon}^4 / \{ \pm1 \})$.

Type(3): increase $b_{4+}$ by 1. The singularity is
$(  B_{\epsilon}^4 / \{ \pm1 \} \times ( B_{\epsilon}^4 / \{ \pm1 \})$.

Type(4A) increase $b_2$ by 1, $b_3$ by 2, $b_{4+}$ by 1, and
$b_{4-}$ by 1.

Type(4B) increase $b_3$ by 2, $b_{4+}$ by 2, and $b_{4-}$ by 2.

The singularity of type(4) is an isometric involution $\sigma$ of
$T^4 \times ( B_{\epsilon}^4 / \{ \pm1 \})$, where $\sigma=( 1/2+
x_1, x_2, -x_3, -x_4, y_1, y_2, -y_3, -y_4)$. Namely, the singular set
is isomorphic to $( T^4 \times ( B_{\epsilon}^4 / \{ \pm1 \}) )/ < \sigma >$.

Type(5A) increase $b_2$ by 1, $b_{4+}$ by 1, and $b_{4-}$ by 1.

Type(5B) increase $b_{4+}$ by 2, and $b_{4-}$ by 2.

The singularity of type(5) is isomorphic to $ (T^4 / \{ \pm1 \} \times
B_{\epsilon}^4 / \{ \pm1 \} )/ < \sigma >$.

As a result, one found the singular set of this orbifold contains
2 type(1), 8 type(2), 64 type(3) and 4 type(4). If we choose to
have $j$ type(4A) and $4-j$ type(4B) and add up all the Betti
numbers in the twisted sectors as well as the untwisted sector, we
have the Joyce's manifolds $Y_j$ with

\eqn\bettispin{ b_2=10+j,\  b_3= 16,\  b_{4+}=109-j,\ b_{4-}=45-j,
\ \ j=0, ...,4} \eqn\A{  \hat{A} = {1 \over 24 } (-1 + b_1 -b_2 +
b_3 + b_{4+} - 2 b_{4-})=1.}

In fact, the 4 type(4) singularities come from 16 $\gamma$-fixed
$T^4$s. Notice that $\alpha \delta$ acts trivially on these $T^4$s
and the group elements $\alpha$, $\beta$, $\alpha \beta$, and
$\beta \delta$ act freely on them and reduce the number of $T^4$s
to be 4. Therefore, we have RR ground states
$|0,0;\tilde{f}=1,2,3,4 \rangle_{\gamma}$ corresponding to the
harmonic two forms of the exceptional divisors. Similarly, the
$\alpha \delta$ parity of $|0,0;\tilde{f} \rangle_{\gamma}$ is
also given by the discrete torsion $\epsilon_{\gamma, \tilde{f}}$.
Since the action of $\alpha \delta$ inverses direction 4 and 7, we
can construct RR ground states accordingly as follows.

For $\alpha \delta$ parity even case, we have, \eqn\spinRRstates{
\eqalign{ &\epsilon_{\gamma; \tilde{f}}=1\ , \cr
&|0,0;\tilde{f}\rangle_{\gamma} \ , \
\psi^{3+}|0,0;\tilde{f}\rangle_{\gamma}\ ,\
\psi^{8+}|0,0;\tilde{f}\rangle_{\gamma}\ ,\
\psi^{3+}\psi^{8+}|0,0;\tilde{f}\rangle_{\gamma}\ , \
\psi^{4+}\psi^{7+}|0,0;\tilde{f}\rangle_{\gamma}\ , \ \cr
&\psi^{3+}\psi^{4+}\psi^{7+}|0,0;\tilde{f}\rangle_{\gamma}\ ,\
\psi^{4+}\psi^{7+}\psi^{8+}|0,0;\tilde{f}\rangle_{\gamma}\ ,\
\psi^{3+}\psi^{4+}\psi^{7+}\psi^{8+}|0,0;\tilde{f}\rangle_{\gamma}\
.}}

For $\alpha \delta$ parity odd case, the RR ground states are,
\eqn\spinRRstatestwo{ \eqalign{ &\epsilon_{\gamma; \tilde{f}}=-1\
, \cr &\psi^{4+}|0,0;\tilde{f}\rangle_{\gamma}\ , \
\psi^{7+}|0,0;\tilde{f}\rangle_{\gamma}\ ,\ \cr
&\psi^{3+}\psi^{4+}|0,0;\tilde{f}\rangle_{\gamma}\ ,\
\psi^{3+}\psi^{7+}|0,0;\tilde{f}\rangle_{\gamma}\ ,\
\psi^{4+}\psi^{8+}|0,0;\tilde{f}\rangle_{\gamma}\ ,\
\psi^{7+}\psi^{8+}|0,0;\tilde{f}\rangle_{\gamma}\ , \ \cr
&\psi^{3+}\psi^{4+}\psi^{8+}|0,0;\tilde{f}\rangle_{\gamma}\ ,\
\psi^{3+}\psi^{7+}\psi^{8+}|0,0;\tilde{f}\rangle_{\gamma}\ . }}

Obviously, we obtain $\Delta b_2=1, \Delta b_3=2, \Delta b_4=2$ in
parity even case, and $\Delta b_3=2,\Delta b_4=4$ in parity odd
case, which agrees with the mathematical analysis in \joyce.

\subsec{spin(7) extended superconformal algebra}

Consider a direct product space $M \times S^1$, where $M$ is a
manifold with $G_2$ holonomy. It is always possible to define a
spin(7) structure. And the Cayley 4-form $\phi_4$ in this manifold
with spin(7) structure can be written as, \eqn\fourform{\phi_4 = *
\phi_3 + \phi_3 \wedge dx^8,} where $\phi_3$ is the calibrated
three form in the $G_2$ manifold.

It is true that in the example in the previous section $T^7 /
\langle \alpha , \gamma , \delta \rangle$ gives rise to a Joyce's
7-manifold of $G_2$ holonomy, if we forget about 8-direction. In
fact, we can have a more generic statement which is $T^7 / \langle
\alpha , \gamma , \delta \rangle$ is $always$ a manifold with
$G_2$ holonomy for any choices of the constants $c_i$ and $d_i$
\acharyafive, \ where

\eqn\spinsub{ \eqalign{ \alpha &= ( -x_1, -x_2, -x_3, - x_4, x_5,
x_6, x_7)\ , \cr \gamma &= ( c_1-x_1, c_2-x_2, x_3, x_4, c_5-x_5,
c_6-x_6, x_7)\ , \cr \delta &= ( c_1-x_1, x_2, c_3-x_3, x_4,
c_5-x_5, x_6, c_7-x_7)\ .}}

If we reformulate the action of $\beta$ in the previous section,
we will find that $\beta$ acts as,

\eqn\ztwo{\beta : x^8 \to - x^8, \ \beta^{*}( \phi_3)=-\phi_3, \
\beta^{*}( * \phi_3)= * \phi_3.}

In this example, $\beta$ indeed turns the spin(7) structure into
the spin(7) holonomy. However, it is not clear that we can always
form manifolds with spin(7) holonomy by modding out this kind of
$Z_2$ involution on $G_2 \times S^1$.

Therefore, at least in Joyce's orbifolds, the relation \fourform \
enables us to write down the expression of the stress energy
tensor $T_{spin(7)}$ and the supercurrent $G_{spin(7)}$ in terms
of the corresponding quantities in $G_2$ manifolds \SV .

\eqn\spingenerator{ \eqalign{ T_{spin(7)} &= T_{G_2} + {1 \over 2}
: \partial x^8 \partial x^8 : - {1 \over 2} : \psi^8 \partial
\psi^8 : \ , \cr G_{spin(7)} &= G_{G_2} + : \psi^8 \partial x^8: \
, \cr X_{spin(7)} &=  X _{G_2} + \Phi_{G_2} \psi^8 + {1 \over 2}
\psi^8
\partial \psi^8 \ ,  \cr M_{spin(7)} &= [ G_{spin(7)}, X_{spin(7)}]
\cr &= \partial x^8 \Phi_{G_2} - K_{G2} \psi^8 - M_{G_2} + {1
\over 2} \partial^2 x^8 \psi^8 - { 1 \over 2} \partial x^8
\partial \psi^8 \ . }}

From these generators for the extended supercomformal algebra, it
is not difficult to see that the combination of the $G_2$
automorphism \mirrorGtwo \ and the T-duality in 8-direction is an
automorphism in the algebra. In addition, the T-duality in
\directiongtwotwo \ is also an automorphism in the algebra.
Therefore, we have a list of 14 T-dualities on $T^4$ toroidal
fibrations which generate the mirror symmetry,

\eqn\direction{ \eqalign{ \{ &(2,4,6,8), (2,3,5,8), (1,2,7,8),
(1,3,6,8),(1,4,5,8),(3,4,7,8),(5,6,7,8), \cr
&(1,2,5,7),(1,4,6,7),(3,4,5,6),(2,4,5,7),(2,3,6,7),(1,2,5,6),(1,2,3,4)
\} \ . }}

The first line consists of T-dualities in directions in
\directiongtwo \ and 8-direction. The second line is the same as
the directions listed in \directiongtwotwo. In this spin(7) case,
we don't have the similar relation like \mirrorGtwo. Therefore, in
order to visualize the automorphism in the algebra, we have to
express the spin(7) generators and the algebra in terms of $G_2$
generators and construct our desirable mirror transformation from
$G_2$ automorphism \mirrorGtwo \directiongtwo \directiongtwotwo.
Finally, the expression of $\alpha \delta$ in $\gamma$-fixed
sector is,

\eqn\AD{ \alpha\delta={1 \over 4} \psi_0^4 \psi_0^7
\tilde{\psi}_0^4 \tilde{\psi}_0^7 \epsilon_{\gamma ; \tilde{f}} \
\ .} By the same reasoning, the 14 T-dualities are divided into
two sets $J_4^{\pm}$. Their action is also summarized as follows.

\eqn\directionspin{ \eqalign{ &(i_1, i_2, i_3, i_4 ) \in J_4^{+}
\cup J_4^{-}\ ,  \cr &J_{4}^{+} = \{
(2,3,5,8),(1,3,6,8),(3,4,7,8),(1,4,6,7),(2,4,5,7),(1,2,5,6) \}\ ,
\cr &J_4^{-} = \{ (2,4,6,8),
(1,2,7,8),(1,4,5,8),(5,6,7,8),(1,2,5,7),(3,4,5,6), \cr &\ \ \ \ \
\ \ \ \ \ (2,3,6,7),(1,2,3,4) \}\ . }}

\eqn\spinsummary{ \eqalign{  IIA(IIB) / Y_j &\longleftrightarrow
IIA(IIB)/ Y_{4-j}, \ \ under \ J_4^{-}\ ,  \cr IIA(IIB) / Y_j
&\longleftrightarrow IIA(IIB)/ Y_j, \ \  under \ J_4^{+}\ .}}


\newsec{Conclusion}
In this paper I have generalized the construction of \Gaberdiel \
to the Joyce's manifolds with spin(7) holonomy and shown how the
mirror symmetry is realized in the superconformal algebra as a
combination of a T-duality in 8-direction and a $G_2$ mirror
symmetry transformation, or a combination of 2 distinct $G_2$
mirror transformations. The spin(7) mirror transformation contains
14 different kinds of T-dualities on the $T^4$ fibrations. By an
analysis on the change of discrete torsion, one can classify these
14 T-dualities into 2 kinds, one of which changes the discrete
torsion and the other does not.

In \keshav, the authors completed a cycle of the dualities by
explicitly performing the T-duality on $T^3$ fibration and a $G_2$
flop in M-theory. It would be interesting to generalize the
computation to a duality cycle involving spin(7) and $G_2$
manifolds and understand how the generalized mirror symmetry lies
in this picture \gukov .

In order to understand the $G_2$/spin(7) mirror symmetry better,
one may try to T-dualize the known various non-compact metric
solutions with $G_2$/spin(7) holonomy \cvetic \ and see how they
are connected through mirror symmetry. In the Calabi-Yau case,
NS-NS fluxes can turn the CY target space into half-flat \Gurrieri
. The generalized mirror symmetry for $G_2$ and spin(7)in the
presence of the background fluxes also begs some further study.
Finally, it would also be interesting to see how we can fit the
$G_2$ or spin(7) mirror symmetry into the correspondence of
heterotic($G_2$)/ IIA($G_2$ orientifold)/M-theory(spin(7))
\majumder .

\centerline{\bf Acknowledgements} I would like to thank Peter
Kaste for his useful comment on the draft. The research of WYC is
supported by supported by the U.S. Department of Energy under
contract number DE-AC03-76SF00515.



\listrefs

\end